\documentclass[traditabstract]{aa} 
\usepackage{graphicx}
\usepackage{txfonts}
\begin{document}
   \title{Pre-nova X-ray observations of V2491\,Cyg (Nova Cyg 2008b)}

   \subtitle{}

   \author{A. Ibarra\inst{1}\thanks{INSA (Ingenier\'ia y Sistemas Aeroespaciales S.A.)},
          E. Kuulkers\inst{1}, 
	  J.P. Osborne\inst{2}, 
	  K. Page\inst{2}, 
	  J.U. Ness\inst{1},
	  R.D. Saxton\inst{1},
	  W. Baumgartner \inst{3,4}
	  V. Beckmann \inst{5,6},
	  M.F. Bode\inst{7},
	  M. Hernanz\inst{8},
	  K. Mukai\inst{9},
          M. Orio\inst{10,11},
	  G. Sala\inst{12},
	  S. Starrfield\inst{13},
	  G.A. Wynn\inst{2}
}

   \institute{ESA, European Space Astronomy Centre (ESAC), Apartado 78, 28691, Villanueva de la Ca\~nada (Madrid), Spain\\
              \email{Aitor.Ibarra@sciops.esa.int}   
	 \and
	     The Department of Physics and Astronomy, The University of Leicester, Leicester, LE1 7RH, UK 
	 \and
	     NASA/Goddard Space Flight Center, Code 662, Greenbelt, MD 20771, USA
         \and
             University of Maryland, Baltimore County, 1000 Hilltop Circle, Baltimore, MD 21250, USA
	 \and
	     ISDC Data Centre for Astrophysics, Chemin d'\'Ecogia 16, 1290 Versoix, Switzerland
	 \and
	     Observatoire Astronimique de l'Universit\'e de Genev\`e, Chemin des Maillettes 51, 1290 Sauverny, Switzerland
	 \and 
	     Astroph.\ Research Institute, Liverpool John Moores Univ., Twelve Quays House, Egerton Wharf, Birkenhead CH41 1LD 7, UK
	 \and
	     IEEC-CSIC, Campus UAB, Facultat Ci\`encies, C5-par, 2on, 08193 Bellaterra, Spain
	 \and
	     CRESST and X-ray Astrophysics Laboratory NASA/GSFC, Greenbelt, MD 20771, USA
	 \and
	     INAF - Osservatorio Astronomico di Padova, vicolo Osservatorio, 5, I-35122 Padova, Italy
	 \and
	     Department of Astronomy, University of Wisconsin, 475 N. Charter Str., Madison WI 53706, USA
	 \and
	    Grup d'Astronomia i Astrof\'isica, Dept.\ F\'isica i Enginyeria Nuclear, Univ.\ Polit\`ecnica de Catalunya, E-08036 Barcelona, Spain
         \and
	     School of Earth \& Space Exploration, Arizona State University, P.O. Box 871404, AZ 85287-1404, USA
}

   \date{Received January 23, 2009; accepted February 23, 2009}

 
  \abstract{Classical novae are phenomena caused by explosive hydrogen
burning on an accreting white dwarf.  So far, only one classical nova
has been identified in X-rays before the actual optical outburst
occurred (V2487\,Oph).  The recently discovered nova, V2491\,Cyg, is
one of the fastest ({\rm He/N}) novae observed so far.  Using archival
{\it ROSAT}, {\it XMM-Newton} and {\it Swift} data, we show that
V2491\,Cyg was a persistent X-ray source during its quiescent time
before the optical outburst. 
We present the X-ray spectral characteristics and derive X-ray fluxes.
The pre-outburst X-ray emission is
variable, and at least in one observation it shows a very soft X-ray source.}


   \maketitle
%

\section{Introduction}
Classical (and recurrent) novae are the third most energetic outbursts
after GRB and SNe, but are much more common than either of those types
of catastrophic explosions. They take place in binary systems in which
mass is transferred from a low-mass secondary star onto a white
dwarf. The accreted material gradually becomes degenerate, and when
temperatures become high enough and the pressure at the bottom of the
accreted envelope is sufficient, a thermonuclear runaway
results. Enough energy is deposited in the accreted material to eject
a fraction of the envelope of the white dwarf (Bode \&\
Evans~\cite{Bode}).

Novae can be classified into two observational groups: {\rm Fe\,{\sc II}} and
{\rm He/N}, respectively, depending on the emission lines detected in
their optical spectra (Williams~\cite{Williams2}). Novae with
prominent {\rm Fe\,{\sc II}} lines evolve more slowly, have lower levels of
ionization, and show P Cygni absorption components. Novae with
stronger lines of {\rm He} and {\rm N} have larger expansion velocities
and a higher level of ionization, and the lines are more  flat-topped
with little absorption. Only a few X-ray observations of {\rm He/N}
novae exist and because of their rapid evolution little is known about
them.

The nova V2491\,Cyg (Nova Cyg 2008b) was discovered on $t_{0}$=2008
April 10.8 UT at about 7.7 mag on unfiltered CCD frames (Nakano et
al.~\cite{Nakano}).  Tomov et al.\ (\cite{Tomov1}) found an optical
V/R/I decline rate from $\sim$0.3 to 0.15 mag per day between $t_{0}+
2.3$ and $t_{0}+ 7.3$ days.  Assuming $t_{0}+ 0.6$ and V=7.54 mag for
the maximum (Nakano et al.~\cite{Nakano}), the time $t_2$ in the V
band is 4.6 days, where $t_2$ is defined as the elapsed time to
decrease 2 magnitudes in its visual luminosity. This $t_2$ value makes
V2491~Cyg a very fast nova, similar to V838\,Her (O'Brien et
al.~\cite{O'Brien}) and V2487~Oph (Hernanz \&\ Sala
\cite{Hernanz}). Optical post-outburst spectra showed V2491\,Cyg to be
a nova in its early phase of outburst (Ayani \&
Matsumoto~\cite{Ayani}, Lynch et al.~\cite{Lynch}).  V2491\,Cyg has
been classified as a He/N nova based on the photometric and
spectroscopic results (Lynch et al.\ \cite{Lynch}, Tomov et al.\
\cite{Tomov2}, Helton et al.\ \cite{Helton}).  The very fast decline
and the optical spectral characteristics, such as extremely broad
lines with complex profiles, and large expansion velocities
($\sim$4000--6000\,km\,s$^{-1}$), mark V2491\,Cyg as a peculiar and
extremely fast nova (Ashok et al.~\cite{Ashok}, Lynch et al.~\cite{Lynch}, Tomov et
al.~\cite{Tomov1}, \cite{Tomov2}).

From optical measurements the reddening towards the source has been
determined: E(B$-$V)=0.3 (Lynch et al.~\cite{Lynch}) and E(B$-$V)=0.43
(Rudy et al.~\cite{Rudy}) on $t_0$+2 and $t_0$+7~days,
respectively. Taking the latter value and using the correlation
between visual extinction, $A_{\rm V}$, and the dust (and hydrogen)
column densities, assuming there is no intrinsic absorption ($N_{\rm
H}$~[cm$^{−2}$]/$A_{\rm V}$ = 1.79$\times$10$^{21}$, e.g., Predehl \&\
Schmitt~\cite{Predehl}), and using $A_{\rm V}$ = 3.1E(B$-$V), the
optical reddening corresponds to $N_{\rm
dust/H}$$\simeq$2.4$\times$10$^{21}$\,cm$^{-2}$. Using the reddening
of E(B$-$V)=0.43 a distance of
10.5 kpc was estimated by Helton et al.~(\cite{Helton}).

\begin{figure}
  \centering
   \includegraphics[height=6.5cm,width=7.cm,clip=]{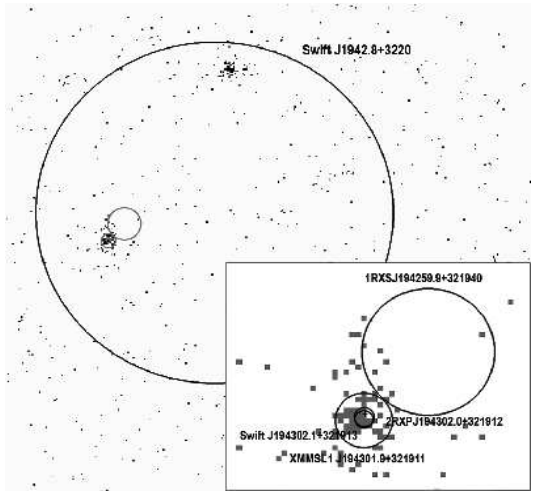}
      \caption{{\it Swift}/XRT X-ray image of the V2491\,Cyg field of view. The big circle corresponds to the {\it Swift}/BAT uncertainty and the zoomed view corresponds to the pre-nova X-ray counterpart uncertainties of Table~\ref{sources}. The cross corresponds to the optical coordinates.}
\label{XRT_Image}
\end{figure}

The object was fainter than 14 mag at $t_0$$-$2 days (Nakano et
al.~\cite{Nakano}).  Balman et al.~(\cite{Balman}) did not detect the
pre-nova down to $R$=18.6$\pm$0.5, 5--9 months prior to the outburst.
However, the pre-nova was detected in archival plates spanning
$\sim$16~years showing a persistent source at R$\simeq$16.3
(Jurdana-Sepic \& Munari \cite{Jurdana}). This indicates a dimming of
the source by about 2~mag several months before the nova outburst.  A
search through archival X-ray data showed the presence of an X-ray
source, before the nova outburst, at the nova position (Ibarra \&\
Kuulkers \cite{Ibarra1}, Ibarra et al.\ \cite{Ibarra2}). We here
describe in more detail the serendipitous X-ray observations performed
by {\it ROSAT}, {\it XMM-Newton} and {\it Swift} of the
pre-nova. Preliminary results have already been reported by Ibarra \&\
Kuulkers (\cite{Ibarra1}) and Ibarra et al.\ (\cite{Ibarra2}).  Since
the discovery, follow-up X-ray observations have been made with {\it
Swift} (Kuulkers et al.~\cite{Kuulkers}, Page et al.~\cite{Page},
Osborne et al.~\cite{Osborne}, Page et al.\ in prep), {\it XMM-Newton}
(Ness et al.~\cite{Ness1}, \cite{Ness2}, in prep.) and {\it Suzaku}
(Takei et al., in prep.).

\section{X-ray observations}

A search through the X-ray catalogues (Available through Vizier at the
Centre de Donn\'ees de Strasbourg (CDS)) showed that the the field of
V2491\,Cyg was observed at different epochs before the nova outburst
by {\it ROSAT}, {\it XMM-Newton} and {\it Swift}. We determine the
Swift position by combining all the data into one single image and
using the {\it ximage} tool.  In Table\ref{sources}, we list the
positions and position errors of the candidate X-ray counterparts. In
Fig.~\ref{XRT_Image}, we show a {\it Swift}/XRT image with all the
pre-nova X-ray counterpart uncertainties. The following paragraphs
describe these satellites and the relevant instruments. The results
are summarised in Table~\ref{Observations}.

The {\it ROSAT} satellite (Tr\"umper et al.~\cite{Trumper}) produced
an all-sky survey (RASS) in the energy range 0.2--2.4 keV and covered
about 25\% of the sky during pointing observations with the PSPC
camera (Voges et al.~\cite{Voges}). Two X-ray sources in these
catalogs have coordinates close to that of V2491\,Cyg, i.e., the RASS
source 1RXS\,J194259.9+321940 and the source 2RXP\,J194302.0+321912
(see also Table~\ref{sources} and Fig.~\ref{XRT_Image}). We have
re-analyzed the 2RXP\,J194302.0+321912 observation using the HEASOFT
(http://heasarc.nasa.gov/docs/software/lheasoft/) software suite.  We
filtered the ROSAT event list with {\sc XSELECT}, using an 80'' source
extraction region and a local background region.

{\it XMM-Newton} (Jansen et al.~\cite{Jansen}) is performing a
sensitive survey of the sky in the 0.2--12\,keV energy band, when it
slews between targets. A catalogue of slew point sources has
been published by Saxton et al.~(\cite{Saxton}). 

\begin{figure}
  \centering
   \includegraphics[height=9.0cm,width=6.cm,angle=-90,clip=]{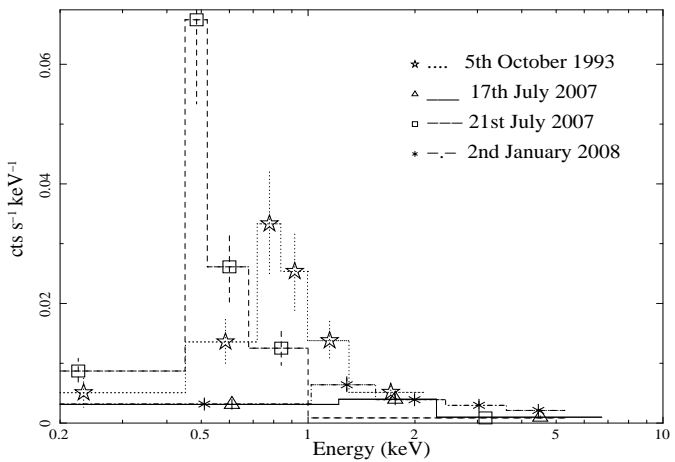}
      \caption{{\it ROSAT}/PSPC (one) and {\it Swift}/XRT (three) spectra of pre-nova X-ray observations. 
Shown are the data points for all the observations.}
\label{Fig_preNova}
\end{figure}

We report on {\it XMM-Newton} observations by the EPIC-pn camera
(Str\"uder et al.~\cite{Struder}) operating in full-frame mode with
the medium filter during a slew performed on 2006 November 15
(OBSID.\ 9127000003). V2491\,Cyg passed through the EPIC-pn field
of view at a large off-axis angle such that an effective on-axis
exposure time of 3.4\,s was achieved. A source with 4 photons was
detected in the full (0.2--12\,keV) energy band of the EPIC-pn with a
significance of 3.5$\sigma$ above the very low background
(0.2 cts/arcmin$^2$) of the slew image. There were no {\it
XMM-Newton} pointed observations of this field before outburst.

Six {\it Swift} observations were made of the V2491\,Cyg field 
before the nova outburst, as part of a follow-up
survey aimed at identifying the X-ray counterpart for sources
discovered by the Burst Alert Telescope (BAT; Barthelmy et
al.~\cite{Barthelmy}). The survey (Tueller et al.~\cite{Tueller};
Tueller et al.\ in prep.) pointing was towards Swift\,J1942.8+3220 
(Sect.~\ref{batSource}).

The {\em Swift} X-ray Telescope (XRT, 0.3--10 keV; Burrows et
al.~\cite{Burrows}) detector was operated in photon counting (PC)
mode, which provides two-dimensional imaging, spectral information,
and 2.5\,s time resolution. All the {\it Swift}/XRT data have been
processed through {\sc xrtpipeline} (Swift\_Rel2.8), with
local-background subtracted spectra binned to
$\ge$20\,cts\,bin$^{-1}$.  One of the XRT observations was too short
(20\,s on 2007 July 15) to be useful, and has not been included in our
spectral analysis.

No optical {\em Swift}/UVOT data were available. This is because the UVOT telescope was
operated with BLOCKED filter due to the presence of a bright
star in the field of view.

\begin{table*}
\caption{Pre-nova X-ray counterpart source identifications, coordinates,
uncertainty, and offset ($\delta$) from the position of the optical nova
(19h43m01.96s, +32d19'13.8"; Nakano et al.~\cite{Nakano}).}
\label{sources}      
\centering          
\begin{tabular}{cccc}     
\hline\hline       
name & position & error & $\delta$ \\ 
     & (RA,Dec; J2000.0)  &  (90\%\/) & \\
\hline                    
1RXS\,J194259.9+321940 & 19h42m59.9s, +32d19'40.5" & 28" & 37"\\ 
2RXP\,J194302.0+321912 & 19h43m02.0s, +32d19'12.0" & 4" & 1.9"\\ 
XMMSL1\,J194301.9+321911 & 19h43m02.0s, +32d19'10.5" & 12" & 3.7"\\
Swift\,J194302.1+321913 & 19h43m02.0s, +32d19'11.0" & 3.7" & 2.8"\\
\hline                  
\end{tabular}
\end{table*}

\section{Results}
\subsection{The pre-nova X-ray identification}

In Table~\ref{sources} we give the positions and source names of the
X-ray sources we identify close to the optical position of V2491\,Cyg.
Based on the {\it Swift}/XRT observations, the pre-nova X-ray source
has been designated as Swift\,J194302.1+321913 (Ibarra et al.\
\cite{Ibarra2}).  The {\it ROSAT}/PSPC source, 2RXP J194302.0+321912,
the {\it XMM-Newton}/EPIC-pn source XMMSL1 J194301.9+321911, and the
{\it Swift}/XRT source Swift\,J194302.1+321913 are all consistent with
the optical nova coordinates (see also Ibarra \&\
Kuulkers~\cite{Ibarra1}, Ibarra et al.~\cite{Ibarra2}).  The {\it
ROSAT}/PSPC source 1RXS J194259.9+321940 is formally not consistent
with the optical position (see Fig.~\ref{XRT_Image}, but due to its
large uncertainty we consider that this is compatible with
2RXP\,J194302.0+321912.  Based on the positional information, as well
as the similar strength of the sources over time (see next
subsection), we suggest that all the above mentioned X-ray sources are
all the same source, and identify it as the X-ray counterpart to the
pre-nova of V2491\,Cyg.

\subsection{Pre-nova X-ray spectral characteristics}

In Table~\ref{Observations}, we list all X-ray pre-nova observations;
for {\it Swift}/XRT we provide the background corrected count rates in
two bands, and the 68\% confidence Bayesian rate errors (Kraft et
al.~\cite{Kraft}) because of the low numbers of counts. The source is
clearly variable. Based on the information in the two energy bands it
changes dramatically in spectral shape between observations on a time
scale down to 4 days. Especially noticeable is the softness of the
source on 2007 July 21 (see also Fig.~1).

\begin{table*}
\caption{Observation log of the {\it ROSAT}, {\it XMM-Newton} and {\it Swift} pre-nova observations. 
Apart from the exposure time the count rates in different energy bands as well as the absorbed fluxes are given
(see text for more details).
For the second ROSAT observation, the {\it XMM-Newton} and the first two {\it Swift} observations, the flux was calculated 
using a power-law with $\Gamma$ between 1.3 and 5.2, and $N_{\rm H}=2.2\times10^{21}$\,cm$^{-2}$. 
}
\label{Observations}      
\centering          
\begin{tabular}{c c c c c c c c c }     
\hline\hline       
Date & Days     & Instrument & Exposure time  & Count rate$^1$ & Count rate    & Count rate    & 0.2--10.0\,keV flux     \\ 
     & before   &          &    (ks)      & (cts/ks)     & (cts/ks)    & (cts/ks)    &  ($10^{-12}$\,erg\,cm$^{-2}$\,s$^{-1}$)    \\ 
     & outburst &          &                &                & 0.3--1.0\,keV  & 1.0--8.0\,keV  &  absorbed                        \\
\hline
   1990-10-19 & 6383 & {\it ROSAT}/PSPC[1RXS]& 0.39   & 28$\pm$11 & ---- & ----                            & 4.5$^{+19.5}_{-1.8}$ \\
   1993-10-05 & 5302 & {\it ROSAT}/PSPC[2RXP]& 4.3    & 22$\pm$3  & 12$\pm$2 & 8$\pm$2                     & 0.3$\pm$0.1 \\  	      			 
   2006-11-15 & 517  & {\it XMM-Newton}/EPIC & 3.43$\times10^{-3}$  & 1173$^{+613}_{-587}$  & ---- & ----  & 2.5$^{+7.0}_{-1.6}$ \\  	      	
   2007-05-25 & 322  & {\it Swift}/XRT       & 1.06  & 21$\pm$5  & 1.9$^{+1.9}_{-1.3}$  & 20$^{+5}_{-4}$   & 5.5$^{+8.5}_{-1.4}$   \\  	
   2007-06-10 & 306  & {\it Swift}/XRT       & 0.97  & 41$\pm$7    & 11.8$^{+4.2}_{-3.6}$ & 31$^{+7}_{-6}$ & 10.7$^{+16.5}_{-2.2}$   \\
   2007-07-17 & 269  & {\it Swift}/XRT       & 3.82  & 23$\pm$3    & 3.5$^{+1.3}_{-1.1}$  & 19$\pm$3       & 1.0$\pm$0.2  \\
   2007-07-21 & 265  & {\it Swift}/XRT       & 4.87  & 37$^{+3}_{-4}$ & 27.9$\pm$3.2 & 9$\pm$2             & 0.8$\pm$0.2 \\  
   2008-01-02 & 100  & {\it Swift}/XRT       & 5.65  & 24$\pm$2    & 3.9$\pm$1.0 & 20$\pm$2                & 1.6$^{+0.5}_{-0.4}$    \\    
\hline                  
\end{tabular}
\note{Energy bands: {\it ROSAT/PSPC}, 0.2--2.0\,keV; {\it XMM-Newton}/EPIC, 0.2--12\,keV; {\it Swift}/XRT, 0.3--10\,keV.}
\end{table*}

The second {\em ROSAT} and last three {\it Swift} observations provide
sufficient counts for a more detailed X-ray spectral analysis. We
first used a power-law model to describe our data and fit them with
{\sc XSPEC} version 11.3. Using this model and leaving $N_{\rm H}$
free in the fits led to poorly constrained values, especially for
$N_{\rm H}$.  To better constrain $N_{\rm H}$ we did a simultaneous
fit to the four individual spectra, using the same value of $N_{\rm
H}$ while leaving the power-law parameters independent. This resulted
in a reasonable fit ($\chi^2_{red}$=1.2 with 10 d.o.f) with $N_{\rm
H}$=2.2$^{+1.3}_{-1.1}\times10^{21}$\,cm$^{-2}$ (90\% statistical
errors are quoted for all parameters). The photon index values are:
3.6$^{+0.7}_{-0.8}$ for the {\it ROSAT} observation and
1.7$^{+0.5}_{-0.4}$, 5.2$^{+1.2}_{-0.7}$, 1.2$\pm$0.4 for the last
three {\it Swift} observations, respectively. Our value of $N_{\rm H}$
is compatible with that
derived from the optical measurements (see Sect.~1).  The softest
spectrum has a very steep power-law index, compared to the other
spectra. Although spectra of novae are known to show many emission
lines (e.g., Mukai \&\ Orio 2005), X-ray spectral fits using other
models (e.g., the optically-thin thermal plasma model VAPEC,
with abundances of C, Fe fixed at 0.3 solar and N fixed at 8
times solar, typical for novae (Nussbaumer et al.~\cite{Nussbaumer},
Ness et al.~\cite{Ness3}) or Bremsstrahlung) did not provide improved
fits.
We also stacked all {\it Swift} observations (excluding the one taken 265 days
before outburst; see Table~\ref{Observations}) to get better
statistics. Again, fits to the resulting spectrum show a slightly
better reproduction of the data when using a power law compared to a
VAPEC model. 
Unfortunately, the quality of the data prevents us from investigating in more detail
multi-temperature emission-line models.

Using the best-fit power-law spectral fits, we derive unabsorbed 
X-ray (0.2--10\,keV) flux values
between 1 and 30$\times$10$^{-12}$\,erg\,cm$^{-2}$\,s$^{-1}$, taking
into account the observations from the various instruments. 

\subsection{The unidentified source Swift\,J1942.8+3220} 
\label{batSource}

It is worth considering whether V2491\,Cyg might be the
counterpart to the {\it Swift}/BAT detection Swift\,J1942.8+3220,
whose error circle encompasses the nova position, see
Fig.~\ref{XRT_Image}. The {\it Swift}/BAT survey catalog source
Swift\,J1942.8+3220 has an estimated average flux of about
2$\times$10$^{-11}$\,erg\,cm$^{-2}$\,s$^{-1}$ (14--195\,keV; Tueller
et al.~\cite{Tueller}). Using public archival {\it
INTEGRAL}/IBIS/ISGRI observations performed between March 2003 and
December 2006 (for a total exposure time of about 620\,ksec) we derive
a $3\sigma$ upper limit (18--50\,keV) of about
1$\times$10$^{-11}$\,erg\,cm$^{-2}$\,s$^{-1}$ at the position of
V2491\,Cyg, more or less comparable with that derived for the {\it
Swift}/BAT source.

Apart from Swift\,J194302.1+321913 there is another XRT source within
the $\sim$5' BAT error circle of Swift\,J1942.8+3220 (see upper part
of Fig.~\ref{XRT_Image}): Swift J194245.9+322411 (RA = 19h42m45.9s,
Dec = +32d24'10.7", with 90\% confidence error radius of 3.6").  This
source has an absorbed AGN-like X-ray spectrum; a power-law fit gives
$\Gamma$=1.5$^{+0.3}_{-0.4}$ and $N_{\rm H}$ =
(2$\pm$1)$\times$10$^{21}$\,cm$^{-2}$ with an unabsorbed 0.3--10\,keV
flux of 1.1$\times$10$^{-12}$\,erg\,cm$^{-2}$\,s$^{-1}$.  Our above
described {\it ROSAT} and {\it XMM-Newton} identifications are not
compatible with this AGN-like XRT source.  The position of this source
was covered by the {\it XMM-Newton} Slew Survey, but no source was
detected with a 2$\sigma$ upper limit of
2$\times$10$^{-12}$\,erg\,cm$^{-2}$\,s$^{-1}$ (0.2--12\,keV), assuming
the AGN-like XRT spectrum.  The AGN-like XRT source is not bright
enough to account for the BAT detection, assuming it is constant, and
so it is not likely to be the X-ray counterpart to the BAT source. If,
on the other hand, we extrapolate the power-law fits of the {\it
Swift}/XRT spectra of the pre\-nova V2491\,Cyg taken on 2007 July 17
and 2008 Jan 2 (i.e., when the source was hard) into the BAT band, we
get fluxes which are more or less consistent with the BAT measured
flux. If, however, the spectrum is assumed to be due to optically thin
high-temperature thermal emission, the extrapolated flux is a few
orders of magnitude lower.  It remains, therefore, uncertain whether
the BAT source is related to the pre-nova.

\section{Discussion}

We find a persistent, but variable, and at least on one occasion very
soft X-ray source present at the position of the nova V2491\,Cyg from
the {\it ROSAT} survey era up to 3 months before the nova outburst. 
This is only the second nova to be detected in X-rays before eruption
(after V2487\,Oph, Hernanz \&\ Sala~\cite{Hernanz}).
A hard spectral component is suggested by {\it Swift}/BAT,
but the association with the pre-nova is not secure. 

V2487\,Oph, also a fast nova ($t_2$$\simeq$6.3~days,
see Hernanz \&\ Sala~\cite{Hernanz}), was suggested to be
a recurrent nova, because of both the rapid
decline in the optical and the presence of a plateau phase during the
decline (Hachisu et al.~\cite{Hachisu}). Pagnotta et
al.\ (\cite{Pagnotta}) discovered a previous outburst of V2487\,Oph in
1900, confirming its recurrent nature. V2491\,Cyg has been suggested to
be a recurrent nova also (Tomov et al.\ 2008b).

Assuming a distance of 10.5\,kpc (Helton et al.~\cite{Helton}) and
using the power law fit results, we derive 0.2--10\,keV X-ray
luminosities ranging from about 1$\times$10$^{34}$ to
4$\times$10$^{35}$\,erg\,s$^{-1}$. This is comparable to that derived
for V2487\,Oph, about 3 years after the outburst, i.e.,
$\simeq$8$\times$10$^{34}$\,erg\,s$^{-1}$ (0.2--10\,keV; Hernanz et
al.~\cite{Hernanz2}) assuming a distance of 10 kpc (but note that the
distance is rather uncertain, see Hernanz \& Sala~\cite{Hernanz}).  It
is, however, orders of magnitude higher than that seen for the
recurrent nova RS\,Oph about 2 years after the 2006 outburst
($\simeq$5$\times$10$^{31}$\,erg\,s$^{-1}$, Nelson \&
Orio~\cite{Nelson}, using a distance of 1.6 kpc, Mason et
al.~\cite{Mason}, Bode et al.~\cite{Bode}). This is compatible with
the fact that in general fast novae appear to be brighter than slow novae, when
they are at quiescence (e.g., Becker \& Marshall~\cite{Becker}, Orio
et al.~\cite{Orio}). 

Our X-ray luminosity estimates imply inter-nova mass accretion rates
in the range $10^{-9}$ - $10^{-8}$ $M_\odot$ yr$^{-1}$, for a 1
$M_\odot$ white dwarf. These rates are around an order of magnitude
lower than those required (judged from the models of Yaron et
al.~\cite{Yaron}) to fuel the novae outbursts of RS Oph, which are
recurrent on a $\sim$ 20 year period. This would imply that, even with
a massive white dwarf, the recurrence timescale for novae in V2491 Cyg
would be $>\sim$ 100 yrs, similar to V2487 Oph.

The soft spectra observed with {\em ROSAT}/PSPC and {\em Swift}/XRT
resemble those of the post-outburst super-soft state (Page et al.\
2009, in prep.), but are much fainter. This, and the fact that the
{\em Swift}/XRT observation was taken less than a year before the nova
outburst, suggests that the spectra were not obtained during a
super-soft state after a previous nova outburst or accretion driven
nuclear burning, in agreement with our relatively low mass transfer
estimates. The {\em Swift}/XRT observations show that V2491\,Cyg
changes spectral state on at least a 4-day time scale.

Magnetic CVs in quiescence show harder spectra than their non-magnetic
equivalents (e.g., Barlow et al.~\cite{Barlow}, Landi et
al.~\cite{Landi}). If the BAT source is the same as that seen at
$\lesssim$10\,keV, V2491\,Cyg may be magnetic. Polars, for example,
are known to change from hard to soft states as the white dwarf
rotates (e.g., Heise et al.~\cite{Heise}), which could explain the
X-ray spectral evolution we observe.  However, we do not see evidence
for short-term ($\sim$hour) orbital related variations. Moreover, in
general polars are rather weak X-ray sources (see King \&\ Watson
\cite{King}). Additionally, at $\lesssim$10\,keV the spectrum is not
unlike that seen in non-magnetic CVs (e.g., Baskill et
al.\cite{Baskill}).  This brings into doubt a magnetic interpretation.


\begin{acknowledgements}
We acknowledge the use of observations obtained with {\em XMM-Newton},
an ESA science mission with instrument and contributions directly
funded by ESA Member States and the USA (NASA), as well as public data
from the {\em Swift} data archive.  JO and KP acknowledge the support
of the STFC; SS acknowledges partial support from NASA and NSF grants
to ASU. We made use of the {\em XMM-Newton} archive tool, ESAC/ESA,
Madrid, Spain. Also, we made use of the VizieR catalogue access tools, CDS,
Strasbourg, France.
\end{acknowledgements}

\end{document}